\documentstyle[aps,prl]{revtex}

\begin{document}
\draft

\twocolumn[\hsize\textwidth\columnwidth\hsize\csname
@twocolumnfalse\endcsname

\title{Rotational Invariance, Phase Relationships and the Quantum Entanglement Illusion}

\author{Caroline H Thompson\cite{CHT/email+}\\
	Department of Computer Science, University of Wales, 	Aberystwyth, \\
	SY23 3DB, U.K.}
\date{\today}
\maketitle

\begin{abstract}
Another Bell test ``loophole" - imperfect rotational invariance - is explored, and novel realist ideas on parametric down-conversion as used in recent ``quantum entanglement" experiments are presented.  The usual quantum theory of entangled systems assumes we have rotational invariance (RI), so that coincidence rates depend on the difference only between detector settings, not on the absolute values.  Bell tests, as such, do not necessarily require RI, but where it fails the presentation of results in the form of coincidence curves can be grossly misleading.  Even if the well-known detection loophole were closed, the visibility of such curves would tell us nothing about the degree of entanglement!  The problem may be especially relevant to recent experiments using ``degenerate type II parametric down-conversion" sources.  Logical analysis of the results of many experiments suggests realist explanations involving some new physics.  The systems may be more nearly deterministic than quantum theory implies.  Whilst this may be to the advantage of those attempting to make use of the so-called ``Bell correlations" in computing, encryption, ``teleportation" etc., it does mean that the systems obey ordinary, not quantum, logic.  

\end{abstract}

\vskip2pc]

\section{Introduction}
Though this paper is largely concerned with general considerations of the role of rotational invariance (RI) in ``EPR" (Einstein-Podolsky-Rosen) experiments, it also introduces some entirely new hypotheses concerning frequency and phase relationships in the degenerate case of parametric down-conversion (PDC).  These are covered mainly in the final part of the paper, starting at section~\ref{Applications}.  Some readers may find the detailed mathematical groundwork of section~\ref{Rotational Invariance} redundant. 
 
There are few references in the literature to the matter of rotational invariance, either in the theory or practice of quantum entanglement.  Early tests -- the key EPR or ``Bell-test" experiments that are commonly accepted as establishing ``quantum nonlocality" as a fact -- involved polarised light from atomic cascade sources~\cite{Freedman72,Aspect81+}.  It seemed reasonable to assume that there was no preferred orientation of polarisation, so that coincidence rates were necessarily dependent on the {\em difference only} between the settings of detectors, not on the particular choice of either setting individually.  Even in these experiments, perhaps more attention should have been paid to testing for invariance (could the polarisation in Aspect's experiments, for example, have been biased towards the vertical -- the polarisation direction of the stimulating lasers~\cite{Aspect83}?), but it is in later experiments, using degenerate PDC sources~\cite{Tittel97,Weihs98,Kwiat98}, that there may have been more serious misinterpretation.  

Admittedly, the actual Bell tests performed may in some cases be general ones, not dependent on rotational invariance, but published papers reinforce the psychological impact of the tests by means of graphs of coincidence rate variations.  In these graphs, one of the detector settings is held fixed.  The high ``visibility" ((max - min)/(max + min)) of such graphs is understood to be corroborative evidence of entanglement.  But what if all the polarisations had in fact been parallel?  High -- even 100\% -- visibility could have arisen from this cause.  It is only if we know that the polarisation direction (or other hidden variable) had RI that the visibility of the curve gains any significance at all (see Appendix~\ref{Loopholes} for yet other realist explanations of high visibility).  It is not usual to conduct comprehensive tests to show that there was no bias towards particular directions, that the graphs would have been just the same had the setting of the ``fixed" detector been chosen differently.  Indeed, in some experiments~\cite{Kwiat98,Kwiat95}, it is clear that the absolute settings have a real physical significance, and this is actively taken into account: the fixed detector is always set at $45^{\circ}$.

Failure of rotational invariance represents effectively yet another ``loophole" in Bell test experiments, and a brief discussion of other common ones is presented in Appendix~\ref{Loopholes}.  Is it not strange, incidentally, to find Bell test violations presented as demonstrating that the logic of the quantum world is different from every-day logic -- that no realist explanation of the results is possible -- for experiments that block only {\em one} of the possible loopholes?  Such a statement would become logical only if {\em all were blocked simultaneously}!

The ``rotational invariance loophole" may be but a minor factor in most Bell test experiments, causing a slight increase in the visibility of the coincidence curve, but it could be an important matter of principle. I shall discuss it with special reference to the experiment by Gregor Weihs {\em et al.} in 1998~\cite{Weihs98}.  Though it is not clear whether or not it was in fact important here -- the relevant information is not available -- the form of the experiment makes it a potentially valuable one for the direct investigation 
of the quantum theory of ``degenerate PDC''.  PDC experiments of this type are currently prominent in the area of ``quantum computing".  It might be as well for those concerned to consider whether or not the purely logical, ``realist", explanation I present might be more robust than the quantum theory (QT) entanglement story.

\section{Rotational Invariance}
\label{Rotational Invariance}
The two extremes for RI are the full RI case generally assumed and the case of total failure, in which one direction is preferred to the exclusion of all others.  There is, however, another special case that may be important, a case that might be termed ``Binary Rotational Invariance", in which the ``hidden variable", instead of taking a continuous range of values, can take only two discrete ones. It can lead to false interpretations of coincidence curve visibilities (though, in all likelihood, beneficial effects so far as practical applications are concerned).

\subsection{Full Rotational Invariance} 
This is the standard case, both under QT and the usual realist model.  It leads to the familiar QT prediction,
\begin{equation}
\label{QT prediction}
P_{ab}^Q = \frac{1}{2}\cos^2(a - b), 
\end{equation}
for the coincidence rate between two `$+$' outcomes, one on each side of the experiment, or the rather less familiar basic local realist one:
\begin{equation}
\label{Furry prediction}
P_{ab}^{BLR} = \frac{1}{8} (1 + 2 \cos^2 (a - b)). 
\end{equation}


\begin{figure}
\begin{center}
\small                          
\input{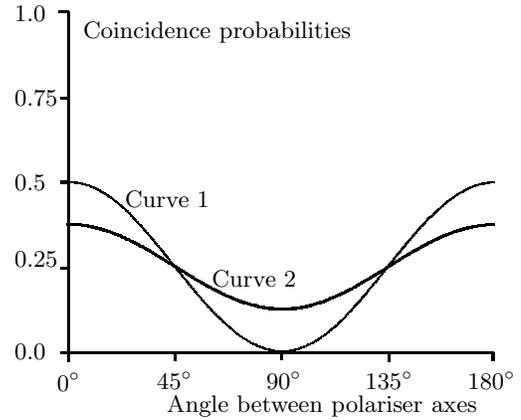}
\end{center}
\caption{The principal predicted ``coincidence curves" for the ideal case.  Curve (1): quantum theory; curve (2): the basic local realist prediction.}
\end{figure}

This last expression comes from just a few straightforward assumptions: 
\begin{enumerate}
\item {\bf ``Locality":}  Once we have fixed the value of a ``hidden variable" that carries all the information on correlation from the source, the actual detections at different localities are independent events.  We can apply the ordinary rules of statistics to calculate the probability of detecting $A$ and $B$ together by multiplying the probabilities for $A$ and $B$ separately.  

\item {\bf Malus' Law:} The detectors are perfect and exactly ``linear", so that Malus' Law, which in its original form tells us that the intensity of the output from a polariser is $\cos^2$ times the input, holds for the probabilities of detection of the light pulses involved in the ``single photon" conditions of the experiments concerned.  Thus we have probabilities of detection given by the formulae:
\begin{eqnarray}
\label{probability for fixed lambda}
f_a^+(\lambda) & = & \cos^2(\lambda - a) \\
& \mbox{for the `$+$' channel and} \nonumber \\
f_a^-(\lambda) & = & \sin^2(\lambda - a) 
\end{eqnarray}
for the `$-$' one, for a pulse of polarisation direction $\lambda$ on exit from a polariser (a Wollaston prism, for example  -- see Appendix~\ref{Wollaston prism}) with axis direction $a$.

\item {\bf Rotational Invariance:} as discussed here.
\end{enumerate}

Both QT and this local realist model predict constant ``singles" rates, the realist calculation being:
\begin{eqnarray}
\label{Singles formula}
p_a^+ & =  & \int^{\pi}_0 d\lambda \frac{1}{\pi} \cos^2 (\lambda - a) \nonumber
 \\
& = & \frac{1}{2}. 
\end{eqnarray}
We simply integrate over all $\lambda$ values, giving each equal weight.  In rather more general situations, the formula would be:
\begin{equation}
p_a^+ = \int^{\pi}_0 d\lambda \rho(\lambda) f_a^+(\lambda),
\end{equation}
where $\rho(\lambda)$ is the density function giving the probability of emission at polarisation angle $\lambda$.

Assuming that polarisations of $A$ and $B$ signals are always parallel, so that values of $\lambda$ are shared, the coincidence rates (for `$++$' coincidences) are calculated by direct integration of:
\begin{equation}
\label{Furry coincidences: basic}
P_{ab}^{BLR} =  \int^{\pi}_0 d\lambda \frac{1}{\pi} \cos^2 (\lambda - a) \cos^2 (\lambda - b),
\end{equation}
where $b$ is the polarisation axis direction of the second polariser.  This, of course, is just a special case of:
\begin{equation}
\label{Furry coincidences: general}
P_{ab}^{BLR} =  \int^{\pi}_0 d\lambda \rho(\lambda) f_a^+(\lambda) f_b^+(\lambda).
\end{equation}

It is worth noting in passing that a subset of the possibilities for the integrand above represents probably the most important realist model of all: that in which the detection loophole alone is responsible for Bell test violations.  All that is needed is to relax the assumption of ``linear" detectors, so that the functions $f$ are not $\cos^2$ (indeed, the $a$ and $b$ ones need not even be the same!).  If the detectors are such that the probability of detection is effectively zero for a {\em whole range of values} near $\lambda - a = \frac{\pi}{2}$, instead of just for one {\em exact point}, the minimum of $P_{ab}^{BLR}$ will be smaller than the value $\frac{1}{8}$ given by the standard formula and the visibility of the curve greater than the 0.5 that it predicts.  It is not hard to construct examples with visibility 1.  This is the ``optical" variant of the ``missing bands" description of the operation of the detection loophole discussed in the author's ``Chaotic Ball" paper~\cite{Thompson96}.

To summarise, the distinguishing features of the ``full RI" case are (a) singles counts that do not vary with detector settings and (b) coincidence counts that are functions of just the difference, $(a - b)$, between detector settings.  As will be shown below, constancy of the singles counts is not sufficient.

\subsection{Complete Rotational Invariance Failure}  
In an atomic cascade experiment using polarised light, complete failure of RI means that all the light is emitted with just one polarisation orientation, vertical, say.  It is easily seen that this means that singles rates are maximum when detectors have their axes vertical; that if one detector is held fixed and the other varied we obtain coincidence curves that all (if Malus' Law is obeyed) have minimum of zero, obtained when the variable detector is horizontal.  The maximum coincidence rate depends on the setting of the ``fixed" detector, but for all settings other than the horizontal one the {\em visibility} is 100\%.  (The fact that visibility depends only on the minimum seems often to be forgotten.  In a regime in which ``normalisation" is customary, this is unfortunate.  A ``true" Bell test, incidentally, would involve unnormalised data only: there are no circumstances in which visibility is a satisfactory substitute.)

If experimental results were to exhibit such behaviour, they would not be taken as demonstrating entanglement, as quite clearly it would be an almost deterministic situation.  Two features of the results would have ruled out entanglement: the full visibility of the singles curves and the strong sensitivity of the coincidence curves to the setting of the fixed detector.  When the fixed detector was horizontal, there would have been no coincidences whatever the setting of the variable one.  In all other cases the coincidence rates would vary but the maximum would depend on the fixed setting.

The formula for predicted coincidences in this ``Complete RI Failure'' case is:
\begin{equation}
\label{Complete RI Failure}
P_{ab}^{CRIF}(\lambda_0) = \cos^2 (\lambda_0 - a) \cos^2 (\lambda_0 - b),
\end{equation}
where $\lambda_0$ is the {\em constant} (vertical, in the above discussion) value of the common hidden variable shared by all the signals.  No integration is really involved in its derivation, though by use of a delta function for the weighting we can artificially frame a derivation in the standard form, following the pattern of equation~(\ref{Furry coincidences: general}).

\subsection{``Binary Rotational Invariance"}
Whilst ``Complete RI Failure" corresponds to changing the distribution of the hidden variable from the constant $\frac{1}{\pi}$ to a delta function centred on one particular value of $\lambda$, the special case to be presented here (that I shall term ``Binary RI") can be obtained by changing it to the sum of two (equal-weight) delta functions, at $\lambda_0$ and $\lambda_1 = \lambda_0 + p/2$, where $p$ is the period involved.  

If the hidden variable is polarisation and the source is an atomic cascade, this model may seem strange, but it possibly fits some PDC situations.  All that it amounts to is that we have 50\% vertically and 50\% horizontally polarised signals.  In many PDC experiments, however, the hidden variable is really, I assert, ``phase difference".  As explained later, there are good reasons to think that this -- in the degenerate case and if there is zero dispersion -- falls logically into two sets, differing by $\pi$ (half the period of $2\pi$).  (Somewhat confusingly, it emerges that mathematically, after projection as in Weihs' 1998 experiment of the two components in a $45^\circ$ direction, we find that this translates into an apparent angular difference of $\frac{\pi}{2}$.  See Appendix~\ref{Wollaston prism}.)

In real experiments dispersion will tend to spread the delta functions out, in extreme cases causing them to merge and eventually blend into the full RI model.  An experiment that could test whether the underlying logic really is as proposed would consist of repeating that of Gregor Weihs {\em et al.}, omitting the random number generation but using the best available filters {\em etc.}~to reduce dispersion to a minimum.  In contrast to the CRIF case, we would expect to obtain singles counts that had negligible variation with detector setting.  Coincidence curves, though, would have similar characteristics to the CRIF ones, their maxima being highly sensitive to the absolute value of the setting of the ``fixed" detector.  

Mathematically, Binary RI means dealing with the average of two equal ensembles, one with $\lambda = \lambda_0 = 0$, say, and the other with $\lambda = \lambda_1= \frac{\pi}{2}$.  

Singles counts for given $\lambda$ are given as usual by:
\begin{equation}
\label{Singles prob, given lambda}
f_a (\lambda) = \cos^2 (\lambda - a)
\end{equation}
(omitting the `$+$' suffix in this and other similar expressions for brevity)
but, with $\lambda$ equalling 0 half the time and $\frac{\pi}{2}$ the other, the observed singles counts averaged over the ensemble will be:
\begin{eqnarray}
\label{Singles prob, BRIF}
p_a^{BRIF} & = & \frac{1}{2} (\cos^2 a + \cos^2 (\frac{\pi}{2} - a)) \nonumber \\
	& = & \frac{1}{2} (\cos^2 a + \sin^2 a) \nonumber \\
	& = & \frac{1}{2}.
\end{eqnarray}

Coincidence counts are predicted by the realist model (integrating expression (\ref{Furry coincidences: general})) to be:
\begin{equation}
\label{Binary RI Failure}
P_{ab}^{BRIF} = \frac{1}{2} (\cos^2 a \cos^2 b + \sin^2 a \sin^2 b),
\end{equation}
which gives full visibility (as $a$ is varied) for $b = 0$ or $\frac{\pi}{2}$ but a constant value of $\frac{1}{4}$ and {\em zero} visibility for $b = \frac{\pi}{4}$.  Clearly the curves depend on the choice of $b$, and this is the distinguishing feature to differentiate between this case and ``full RI". 

\section{Applications to Real Experiments}
\label{Applications}

\subsection{Introduction}

Real experiments using the degenerate PDC sources under consideration are sometimes complicated, and realist models must reflect this logically.  The experiments may not only have hidden variables that are partly binary and partly continuous, but may also suffer from detectors whose characteristics vary with their absolute setting!  In some of Paul Kwiat's recent experiments, for example~\cite{Kwiat98,Kwiat95}, the signals are analysed by physically rotating waveplates~\cite{waveplates} inserted in front of polarising prisms.  This means, I believe, that subensembles of quite different natures will be detected for different settings of the waveplates.  

\begin{figure}
      \small                          
	\input{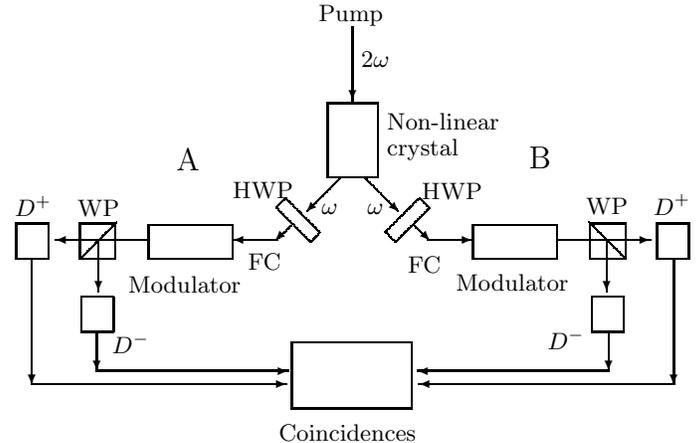}
\normalsize
\caption{Scheme of Weihs' 1998 experiment, omitting random number generation etc.  HWP = half-wave plate and compensating crystal; FC = 500m fibre cable; WP = Wollaston prism; $D^+$, $D^-$ are photodetectors; the box marked ``Coincidences" stands for the complete system that puts time-tags on the results, stores them, then later analyses.}
\label{Weihs' 1998 Experiment}
\end{figure}

For simplicity, therefore, we consider just experiments in which there is no physical rotation.  Weihs' 1998 experiment (Fig.~\ref{Weihs' 1998 Experiment}) is of this kind, and as a further simplification we assume initially that it is dispersion-free: the pump laser operates at just one exact frequency ($2\omega$) and induces ``downconverted" signals of exactly half that frequency.  If the mechanisms controlling the random settings are ignored, the design reduces to (almost) that of the standard EPR experiment.  The fact that the settings are random is of no consequence to the current discussion (see Appendix~\ref{Loopholes}).  The major difference between this and the atomic cascade-type experiments is the nature of the hidden variable, as I shall explain.

We are dealing with ``degenerate type-II parametric down-conversion", in which QT assumes that two simultaneous ``photons" are emitted, one vertically and the other horizontally polarised.  The vertically and horizontally polarised light is emitted in the particular case under consideration in two cones, and the light selected for the ``EPR" experiment is taken from the two lines of intersection~\cite{Kwiat95}.  According to QT, the down-conversion process necessarily converts one pump photon into exactly two output photons, so it is necessarily assumed that we get one vertical and one horizontal and that, when  there are detections in both EPR channels at the same time, one must come from a vertical and one from a horizontal signal.  

But if this were so, how would the ``modulator" produce the results it does?  It functions by using a voltage change to alter the relative speeds of the vertical and horizontal components of light, and hence the alters the relative phase (see footnote 13 of ref.~\cite{Weihs98}).  If each light pulse (I use this expression in preference to ``photon") is polarised in just one direction, then how can the relative phase have any consequence?  Yet the outcome of the experiment is that the modulator setting most certainly does have effect, and I attempt to clarify how this is detected in Appendix~\ref{Wollaston prism}, which discusses the role of the Wollaston prism.  Although the two components are orthogonal, the effect of the prism is to look at projections of them onto $45^\circ$ planes, which enables interference patterns to be formed, much as in other interferometers.  The phase shift of the modulator -- the consequence of a voltage change -- plays the role of an interferometer phase shift.

Thus the vital difference from the QT description is, I should like to suggest, that in reality both components are present in every pulse, or, at least, in the majority of pulses that are detected as coincidences.  The experiment achieves sensitivity to the modulator by enabling the projections to interfere at the prism.

The reader interested in the QT description of events is refered to the vast literature on the ``singlet state" and EPR experiments, and to a brief comment in Appendix~\ref{QT story}.  There are some points of similarity with the realist description, but major logical differences.

\subsection{General considerations re coherence and phases in degenerate PDC}

First, let us take a fresh look at PDC.  I make the fundamental assumption that, in the degenerate case, each signal (pulse of electromagnetic oscillations) is initiated with a particular phase-relationship to the {\em pump} phase, as well as a particular polarisation.  Under QT, this is not possible, as the frequencies are assumed even in the degenerate case to be ``conjugate"~\cite{Conjugate outputs}, not identical.  Indeed, under QT the whole picture of laser light is so different that those trained in the discipline may have difficulty following the classical ideas that seem to me to fit the facts so well.  Perhaps I need to go further back, and explain that I am suggesting that the kind of laser light used in these experiments comprises a series of (long) pulses, each of one pure frequency.  I see no other way to explain the observed interference effects, which imply long coherence lengths, in the original, classical, meaning of the term: the path length difference that can be introduced to a split beam such that recombination will result in interference.

As there is always some dispersion (observed spectra cover a band of frequencies), direct verification of my model is not easy, yet I have been unable to find any experimental evidence to support the QT ``coherent state" concept~\cite{Furusawa98}.  In my model, the dispersion of the (degenerate) down-converted light is merely a matter of its comprising a series of pulses, each of pure frequency, inherited from pump laser light with similar properties\cite{Kwiat/Chiao91}.  Given a pump frequency, $2\omega$ (which may in general vary between pulses), no conservation laws are infringed if the outputs induced by a particular pulse both have frequency exactly $\omega$.  

The phase relationships with the pump arise from the assumption that some causal mechanism is at work.  In some sense, the outputs, with beats half as frequent as the pump, are able to be in phase with every alternate beat, and, since all beats are presumably equivalent, this means a natural 50-50 division into two sets, the ``even" and the ``odd".  (The choice of even or odd might be attributed to the influence of the zero point field, as in the theory of Stochastic Electrodynamics~\cite{Marshall88}.)  The original inspiration for this idea was as an explanation for the ``induced coherence" experiments~\cite{Zou/Wang/Mandel91}.  

The factor of importance for experiments in which both vertical and horizontal components are present in the same pulse is whether their phase sets are the same (giving phase difference zero) or opposite (phase difference $\pi$).  It is not clear whether or not the two components are always generated independently.  In cases in which they are not, it is possible that all phase differences will be zero, but otherwise we would expect the phase differences to be 50\% zero, 50\% $\pi$.  

\subsection{A New Realist Description of Weihs' 1998 experiment}

Returning now to Weihs' experiment, let us follow the progress of an individual ``double" (V and H component) pulse on one side of the apparatus.  We are, for the moment, ignoring dispersion, assuming that the two components are emitted exactly in phase, and treating the initial half-wave plate and the modulator as a single device whose effect is to modify the phase difference.  

We write the equation for oscillations of the electric vector of our pulse in its initial state as:
\begin{equation}
\label{single pulse initial}
\psi(x) = \frac{1}{\sqrt{2}} (\boldmath{i} \cos x + \boldmath{j} \cos x),
\end{equation}
where $x = \omega t$ and $\boldmath{i}$ and $\boldmath{j}$ are unit vectors in the vertical and horizontal directions, as defined by the axes of the nonlinear crystal.  The expression has been normalised to a wave of amplitude and intensity 1 unit.

After passage through the waveplate and modulator, it becomes:
\begin{equation}
\label{single pulse after modulator}
\psi_a(x) = \frac{1}{\sqrt{2}} (\boldmath{i} \cos x + \boldmath{j} \cos (x + 2a)),
\end{equation}
where $2a$ is the induced phase difference, the common phase shift having, with no loss of generality, been ignored.  (More care is needed when we re-introduce frequency dispersion.  $a$ is then seen to be a function of frequency, since the speed of light in the apparatus varies with frequency as well as with polarisation direction.)

We now encounter a Wollaston prism, set at $45^\circ$.  Assuming Malus' law to apply, the effect of the prism is to output the sum of the projections of the waves.  The mathematics is given in Appendix~(\ref{Wollaston prism}), and leads to the results:

\begin{eqnarray}
\label{single pulse probabilities}
f_a^+(0) & = & \cos^2 a \nonumber \\
f_a^-(0) & = & \sin^2 a.
\end{eqnarray}

If all pulses were as above, with initial phase difference of zero (hidden variable $\lambda = 0$), these patterns should appear in the `$+$' and `$-$' channels as we vary the modulator voltage.  But they do not.  The singles counts, we are told, have no oscillations.  This is one strong indicator that what we in fact see could be the superposition of one pattern of this class and one totally out of phase, with $a$ replaced by $a + \frac{\pi}{2}$ (i.e. $\lambda = \frac{\pi}{2}$).  The two patterns could be washing each other out, producing a steady probability averaging 
\begin{equation}
\frac{1}{2}(\cos^2 a + \sin^2 a) \equiv \frac{1}{2}.
\end{equation}

Now to consider coincidences, though.  Do these oscillate?  The answer is ``Yes", and, at least for the values of $b$ selected in Weihs' reported results, the visibility as $a$ is varied is high.  This is the second indicator that what we are seeing is likely to contain a strong element of the ``Binary Rotational Invariance" situation described above.  If it were a pure case (which I assert might happen if frequency dispersion were negligible), the formula for coincidences would be precisely equation~(\ref{Binary RI Failure}) above. 

Thus the realist prediction for coincidences would be:
\begin{equation}
P_{ab}^R = \frac{1}{2} (\cos^2 a \cos^2 b + \sin^2 a \sin^2 b).
\end{equation}
Let us compare this with the standard QT prediction,
\begin{equation}
P_{ab}^Q = \frac{1}{2} \cos^2 (a - b).
\end{equation}

Expanding the latter, we find that it can be written:
\begin{eqnarray}
P_{ab}^Q & = & \frac{1}{2} (\cos a \cos b + \sin a \sin b)^2 \nonumber\\
  	& \equiv &  P_{ab}^R + \cos a \cos b \sin a \sin b.
\end{eqnarray}

The two formulae are identical when the second term is zero, which may explain some of Weihs' results.  

There are problems, though, in attempting a detailed reconciliation between my model and the actual experiment.  Insufficient information is given in the published paper~\cite{Weihs98 info}.   Correspondence with Weihs, though very helpful, has shown that more experimentation is needed.  The differences between the rival models concern the positions of the maxima and minima and the way in which the range of variation changes as we alter the ``fixed" setting, $b$.  The available information does not seem to cover either of these points reliably.

As mentioned earlier, frequency variations will blur the issue.  They give rise to a second, continuous, component that is added to our hidden variable, making its distribution bimodal.  This second component is the phase difference caused by variations in the pump frequency, and it can cause the Binary RI model to merge into the standard realist one, with complete rotational invariance. There are, however, sufficient clues in this and other published papers to suggest that further experimentation would be rewarding.  One of the predictions of the class of models under consideration is the production in certain circumstances of skewed coincidence curves, not pure sinusoidal ones.  Is there a hint of this, perhaps, in Fig.~3 of~\cite{Tittel97-1}?

Note that there are other facets of these experiments that deserve attention, and, indeed, it may be impossible to deduce the true facts without a comprehensive study of all of them.  For example, realist models predict that there would (if the electronics allowed!) be some simultaneous outputs from the `$+$' and `$-$' channels of a single prism, and that the shape of the response curve of the detectors as light intensity is varied plays a critical role.  The necessary investigations would entail direct challenges to the photon model of light and require critical re-evaluation of much experimental evidence. 

\section{Conclusion}

It is entirely possible that the kind of entanglement we see here is, as several authors have said ~\cite{Zeilinger99,Hardy99} no more than a change in our state of knowledge.  It does not involve ``nonlocality", but nor does it really obey the tenets of QT.  It is merely a matter of using information (phase-differences) on one set of entities to select those of another identical set with certain properties.  The process is possible because the correlations are real and strong, involving phase, frequency {\em and} polarisation.  Two other factors almost certainly enhance it: nonlinearity of detectors, making the detection loophole large, and failure of rotational invariance.  Both factors increase the visibility of coincidence curves.  Contrary to widespread belief, high visibility (over 50\%) does {\em not} in itself conflict with local realism.

The ideas of this paper would appear to be easily testable, and might remove all trace of mystery from Weihs' and certain other experiments.  For some experiments, another interesting possibility may need to be taken into account: do some detectors integrate over time in such a way that ``interference" can appear to occur between pulses separated in time~\cite{Chalmers99,Kim99}?

A new area of physics awaits exploration.  How do the exactly-matched (but not entangled!) PDC pairs arise?  A theory is needed that recognises that coincidence measurements (usually) involve individual pairs, with no time-averaging over ensembles except at a later stage.  It needs to recognise that measured spectra tell us only time-averages.  The rules that relate bandwidths to time-intervals are not necessarily relevant when we look at the individual pulse, which may have a much longer coherence length than is currently assumed.  

There are theories to cover the general case of parametric down-conversion, but these predict that frequency variations will always be oppositely correlated, never identical.  They fail to predict clearly the existence of the two phase classes of the degenerate case.  The experiments that might now become transparent include those demonstrating ``induced coherence''~\cite{Zou/Wang/Mandel91}, ``quantum erasure''~\cite{Herzog/Kwiat/Weinfurter/Zeilinger95} and ``teleportation''~\cite{Bouwmeester97}.  

\appendix
\section{Other Bell Test ``Loopholes"}
\label{Loopholes}

The term ``loophole" is a euphemism for the statement that the test applied is not valid as a discriminator between QT and realism as it depends on assumptions that are not universally accepted.  (The tests used in practice are, in John Bell's own words (page 60 of ref.~\cite{Bell87}) ``more or less ad hoc extrapolations" of his original, which would have been valid but is impractical~\cite{Clauser/Horne74,Santos91}.)  The most important loopholes are:
\begin{enumerate}
\item {\bf The ``detection" loophole:}  An interesting paper by members of the quantum optics team in Geneva~\cite{Gisin99-1} has drawn attention to the fact that this it is not logical to neglect this.  Clauser and Horne, in their paper of 1974~\cite{Clauser/Horne74} derived practical Bell tests that do not depend on it, but these have rarely been used.  

Some important papers describing the operation of the detection loophole are refs.~\cite{Thompson96,Pearle70,Marshall/Santos/Selleri83}.  In brief, it is only under quantum theory assumptions that we expect the total population of ``coincidences", involving all four combinations of `$+$' and `$-$' outputs on the two sides, to be a fair sample of the population of emitted pairs.  If this population can vary with detector setting, then clearly an estimate of probability of detection will be biased if it uses (as the standard formula does) the total number of coincidences as divisor.  In realist theories, it is the exception rather than the rule for the sample to be fair in this sense.  If light is a purely wave phenomenon, it is possible for photodetectors to respond nonlinearly to intensity.  This, as is clear from the structure of the basic realist prediction~(\ref{Furry coincidences: general}), can cause high coincidence curve visibilities and associated infringements of Bell tests, but the sample will not have been ``fair".  

It is not usual for experimenters to attempt a direct test for linearity~\cite{Weihs99}.  A test that is sometimes performed is to check that the total of the coincidence counts entering into the estimated ``correlation" is constant.  Failures of constancy may be masked by experimental error, however, and, in addition, the parameter values (zero and $45^{\circ}$) most sensitive to this test are not usually explored.

\item {\bf Subtraction of ``accidentals":}  Adjustment of the data by subtraction of ``accidentals" biases Bell tests in favour of quantum theory.  It is now recognised as not legitimate~\cite{Tittel98-1}, but the reader should be aware that it invalidates many published results~\cite{Thompson99}.

\item {\bf Synchronisation problems:}  There is reason to think that in a few experiments bias could be caused when the coincidence window is shorter than the some of the light pulses involved~\cite{Thompson97}.  These few include one of historical importance -- that of Freedman and Clauser, in 1972~\cite{Freedman/Clauser72} -- which uses a test not sullied by either of the above possibilities.
\end{enumerate}

A loophole that is notably absent from the above list is the so-called ``locality loophole", whereby some mysterious unspecified mechanism is taken as conveying additional information between the two detectors so as to increase their correlation above the classical limit.  In the view of many realists, this has never been a contender.  John Bell supported Aspect's investigation of it (see page 109 of ref.~\cite{Bell87}) and had some active involvement with the work, being on the examining board for Aspect's PhD.  Gregor Weihs improved upon the work in his experiment of 1998~\cite{Weihs98}, but nobody has ever put forward plausible ideas for the mechanism.   Its properties must be quite extraordinary, as it is required to explain ``entanglement" in a great variety of geometrical setups, including over a distance of several kilometers in the Geneva experiments of 
1997-8~\cite{Tittel97,Tittel98-1}.

There may well be other loopholes.  Vladimir Nuri~\cite{Nuri99} is currently studying the possible consequences of the usual experimental arrangement, in which simultaneous `$+$' and `$-$' counts from both outputs of a polariser can never occur as the electronics records only one or the other.  Under QT, they will not occur anyway, but under a wave theory the suppression of these counts will cause even the basic realist prediction (\ref{Furry prediction}) to yield ``unfair sampling''.  The effect is negligible if the detection efficiencies are low, however.

\section{The effect of a Wollaston prism}
\label{Wollaston prism}

A Wollaston prism is a ``2-channel polariser".  If a plane-polarised wave is input, the output intensities are given, in the absence of losses, by Malus' Law.  In other words, it is the projections onto the prism axis and the direction orthogonal to this that are output.

Now in the experiment in question, the prism is set at the fixed angle of $45^\circ$, half way between the polarisation directions of our two components.  Assuming that the two components add linearly, the polariser combines the two projections, each of which will have amplitude $1/\sqrt{2}$ times its input amplitude.  As with any other case of interference, the relative phase of the two components has a striking effect on the outcome.  If there are no losses, the wave function~(\ref{single pulse after modulator}), modelling a single (two-component) wave pulse, will lead to outputs:
\begin{eqnarray}
\label{single pulse after prism}
\psi_a^+(x) & = & \frac{1}{2} (\cos x + \cos (x + 2a))\\ 
& \mbox{and} \nonumber \\
\psi_a^-(x) & = & \frac{1}{2} (\cos x - \cos (x + 2a)).
\end{eqnarray}

Using the trigonometric identity 
\begin{displaymath}
\cos(A + B) + \cos(A - B) \equiv 2 \cos A \cos B,
\end{displaymath}

$\psi_a^+(x)$ can be written:
\begin{equation}
\psi_a^+(x)  =  \cos (x + a) \cos a,
\end{equation}
with a similar expression for $\psi_a^-(x)$. 

This is a plane wave whose amplitude is $\cos a$.  Effectively, it represents the interference pattern between the vertical and horizontal components of a single pulse.  If our detectors are linear (detection rates proportional to intensities -- an assumption that can be challenged~\cite{Thompson99}) and all pulses identical, the singles count rates should follow the pattern:

\begin{equation}
\label{single pulse detection rate}
p_a^+  =  \cos^2 a.
\end{equation}

The pulse in question, with zero phase difference apart from that induced by the apparatus, is to be considered as having ``hidden variable" of zero, so that we have $f_a^+(0) = \cos^2 a$.

\section{Note on QT Description of Weihs' Experiment}
\label{QT story}
Weihs expresses the ``entangled state" of the two photons (after their passage through a half-wave plate~\cite{waveplates}) as follows: 
\begin{equation}
\label{entangled}
|\Psi\rangle = 1/\sqrt{2}(|H\rangle_1|V\rangle_2 + 
e^{i\phi}|V\rangle_1|H\rangle_2).
\end{equation}
Apart from the factor $e^{i\phi}$, this is the same ``singlet state" formula as would apply in experiments such as Aspect's~\cite{Aspect81+}.  Whereas for Aspect's polarised light, however, the absence of an experimentally meaningful definition for vertical and horizontal directions means that the formula makes little sense, in the current experiment we can relate it quite closely to realism.  Vertical and horizontal are well-defined, related to the optical axes of all parts of the apparatus, and a direct realist explanation is almost possible.  There are subtle differences, though, so I shall develop realist expressions working from first principles rather then try and force a close correspondence.

One difference involves the $e^{i\phi}$ factor.  Weihs states that he is able to control the phase $\phi$, setting it to $\pi$.  But what is this phase shift?  Under the classical model I am proposing, the only phase difference that is of experimental consequence is the difference between the phases of the two components, when both are present.  We can indeed change this, by means of waveplates or variable ``modulators''~\cite{waveplates}, but the phase difference between the two possibilities represented by the two terms of the wave function~(\ref{entangled}) is, I assert, fixed naturally.  The formula implies that we get {\em either} ($|H\rangle_1$ and $|V\rangle_2$) {\em or} ($|V\rangle_1$ and $|H\rangle_2$), with the ability to adjust the phase between the first and second possibility.  This relative phase is meaningless, as it is between events happening at different times.  Hopefully, the purely realist treatment in the main text clarifies this.


\begin{thebibliography}{00}
\bibitem{waveplates}
Waveplates (and the ``modulator'' used in Weihs' experiment) change the relative phases of the vertical and horizontal components of light.  Ordinary waveplates do this by using birefringent materials, transmitting the two components at different speeds.  The precise speeds vary with frequency, so the resulting phase shifts will do so also.  They do not, incidentally, produce any real ``polarisation rotation'', and this fact is recognised in footnote [13] of Weihs' paper.  When light is analysed by a Wollaston prism after passage through a waveplate, it behaves very much as if its plane of polarisation had been rotated, but subtle tests should be able to reveal the distinction.  Light that appears to have been rotated through $45^\circ$ has in fact been changed from linear polarisation (phase difference $0$ or $\pi$) to circular (phase difference $\frac{\pi}{2}$).
\bibitem{Weihs98 info}
The main data, available electronically, concerns just two settings of each modulator.  Coincidence curves involve further data, but though one modulator takes a continuous range of settings the other is restricted to just the two.  This is insufficient to establish the extent of any failure of RI.
\end{thebibliography}
\end{document}